\documentclass[aps,twocolumn,amsmath,amssymb,preprintnumbers]{revtex4}
\usepackage{amsmath} \usepackage{amsfonts} \usepackage{amssymb}
\usepackage{titlesec}
\usepackage{bbm}
\usepackage{epsfig}
\usepackage{graphics}
\usepackage{graphicx}
 \usepackage[utf8]{inputenc}
\newcommand{\be}{\begin{equation}}
\newcommand{\ee}{\end{equation}}
\newcommand{\ba}{\begin{eqnarray}}
\newcommand{\ea}{\end{eqnarray}}
\newcommand{\nn}{\nonumber}
\newcommand{\kr}{\rangle}
\newcommand{\kl}{\langle}
\newcommand{\h}{{\cal H}}
\newcommand {\mn}{\mu\nu}
\newcommand{\rt}{\rho\tau}
\newcommand {\gr}{_{\rm grav}}
\newcommand {\pf}{\partial F}
\newcommand {\pc}{\partial \chi}

\newcommand {\p}{\partial_\eta}
\newcommand {\D}{{\cal D}}

\titleformat{\subsection}[block]{\normalfont\bfseries}{\thesubsection.}{1ex}{}
\titlespacing{\subsection}{0pt}{10pt}{1pt}[0pt]
\titleformat*{\section}{\large\bfseries}
\renewcommand{\thesubsection}{\arabic{subsection}}

\begin{document}

\title[ ]{Primordial cosmic fluctuations for variable gravity}

\author{C. Wetterich}
\affiliation{Institut  f\"ur Theoretische Physik\\
Universit\"at Heidelberg\\
Philosophenweg 16, D-69120 Heidelberg}

\begin{abstract}

The observability of primordial cosmic fluctuations does not require a geometric horizon $H^{-1}$, which is exceeded temporarily by the wavelength of fluctuations. The primordial information can be protected against later thermal washout even if all relevant wavelengths remain smaller than $H^{-1}$. This is demonstrated by formulating the equations governing the cosmic fluctuations in a form that is manifestly invariant under conformal field transformations of the metric. Beyond the field equations this holds for the defining equation for the correlation function, as expressed by the inverse of the second functional derivative of the quantum effective action. An observable almost scale invariant spectrum does not need an expanding geometry. For a variable Planck mass it can even arise in flat Minkowski space.

\end{abstract}

\maketitle

\section{Introduction}

Indirect observation of the primordial cosmic fluctuations through the cosmic microwave background (CMB) anisotropies is a key for understanding and testing ideas about the ``beginning'' of our universe. In inflationary cosmology, the possibility to receive signals from the beginning, that are not washed out by the long thermal equilibrium period during radiation domination, is usually explained in terms of a horizon. During inflation, the wavelength of fluctuations exceeds the horizon at some given moment (``horizon crossing''). While outside the horizon, the fluctuations are frozen since no signals can be transmitted that affect wavelengths outside the horizon. The information about the beginning of the universe therefore remains preserved. Only much later, typically during the matter dominated epoch, the horizon has grown large enough such that the primordial fluctuations are again affected by physical processes and can be observed. 

The abstract notion of a horizon involves the entire pre-history of a given epoch. The effective freezing of fluctuations is, however, rather determined by the properties at a given time. The above argument can be made more precise if we compare the increase of the wavelength of a fluctuation, $d(a/k)$, with the distance that light travels during the same laps of time, as given by $ad\eta$. Equating these two yields $(1/k)da/a=d\eta$, or for infinitesimal $d\eta,\h=k$. Here $a$ is the scale factor, $k$ the comoving wave number, $\eta$ conformal time, and $\h=\partial\ln a/\partial\eta=Ha$ the conformal Hubble parameter. If light is too ``slow'' as compared to the expansion of the wave length the fluctuation can no longer be modified. One concludes that freezing occurs for the range of modes
\be\label{I1}
k\ll\h.
\ee
The appropriate ``horizon'', to which the wavelength $a/k$ has to be compared, is therefore the inverse Hubble parameter $H^{-1}$. In this discussion the mechanism for preserving information from the beginning of the universe is closely related to the expansion of the universe. It reflects the detailed expansion history of geometry in the inflationary and later epochs. 

We argue in this paper that in models of variable gravity no geometrical expansion is necessary in order to explain the preservation of memory about primordial fluctuations. If the Planck mass is given by a scalar field $\chi$, which is allowed to vary, a freezing of  modes and preservation of primordial information can occur even in the limit of an infinite horizon, $H^{-1}\to\infty$. The basic reason is simple. In our above argument the wavelength $a/k$ should be replaced by an ``observable quantity'' which needs to be dimensionless. For graviton fluctuations this is the wavelength in units of the Planck length, $a\chi/k$. Repeating the same argument we see that freezing of fluctuations occurs for 
\be\label{I2}
k\ll \hat \h,
\ee
with
\be\label{I3}
\hat \h=\frac{d\ln (a\chi)}{d\eta}
\ee
the logarithmic conformal growth rate of the scale factor in units of the Planck length. In other words, modes are frozen once
\be\label{I4}
\frac ka\ll H+\frac{\dot\chi}{\chi}.
\ee

One can construct realistic models of variable gravity \cite{CWVG,CWCG} where $H$ vanishes or is very small. Then fluctuations are frozen once the physical momentum $k/a$ becomes smaller than the logarithmic growth rate of the Planck mass. No geometrical horizon $H^{-1}$ is involved. What is usually attributed to an expanding geometry is now realized by a shrinking Planck length. Similar remarks extend to other quantities. A scale free spectrum will be realized for a constant value of $H+\dot\chi/\chi$. For an appropriately increasing variable Planck mass it is compatible with a static or even shrinking geometry.

Different pictures describing the same physical situation can be related by a conformal rescaling which multiplies the metric by a function of the scalar field $\chi$. ``Field relativity'' \cite{CWVG,VG30} states that physical observables cannot depend on the choice of fields. Different pictures related by conformal rescaling of the metric are therefore completely equivalent. On the level of classical field equations the mapping between two ``frames'' by a conformal rescaling or a ``Weyl transformation'' has been established very early in general relativity \cite{VG123,VG43,VG44,VG45,VG2}. Based on the notion of the quantum effective action it has been argued \cite{VG3} that the equivalence of frames holds for the full quantum field theory. It is sufficient that all observables are expressed as functionals of the effective action and its derivatives. Of course, quantities as temperature etc. have to be rescaled appropriately, such that the dimensionless observable quantities as temperature over mass are the same in all frames. 

Many detailed calculations based on the cosmological field equations have shown the equivalence of frames related by conformal transformations \cite{VG3,VG4,VG117,VG101,VG118,VG119,VG115,VG116,CWEU,JK}. In this spirit, primordial fluctuations in variable gravity models have been discussed in refs. \cite{Pi03,Pi,Liu,Pi11,TQ,ML1,DS,ML2,IS,FLM}, where it has been observed that a scale invariant spectrum is compatible with slow expansion. 

In the present paper we discuss the time evolution of fields and correlation functions in terms of manifestly frame invariant variables. This makes it very clear that primordial fluctuations can be described equivalently in all frames, a property used in practice since a long time (see, for example, ref. \cite{VG71}.) In the usual treatment of fluctuations \cite{Sta1,MUK,Ru,STA,GP,BST,AW} one studies the solution of field equations for small deviations from a homogeneous and isotropic cosmological background. For this part the conformal mapping between different frames is straightforward. 

The linearized equations constitute, however, a system of linear differential equations, such that the amplitude is fixed only once initial conditions are specified. This is done by postulating an initial ``vacuum state'', as the Bunch-Davies vacuum \cite{BD}. The conformal mapping between different states in the quantum mechanics of fields is not much explored and by far not as simple as the  mapping of variables for differential equations. (Non-linear field transformations in a functional integral involve a Jacobian that is often not accessible for practical computations.) We bypass this problem by investigating directly the defining equation for the Green's function \cite{CWneu,CW2}. This allows us to express fluctuation observables directly in terms of the quantum effective action. On this level non-linear field transformations are straightforward. We will be able to formulate the fluctuation problem directly in terms of frame invariant variables.

The power spectrum of fluctuations can be extracted from the (two-point-) correlation function (Green's function, propagator)
\be\label{I5}
G(x,y)=\kl\zeta'(x)\zeta'(y)\kr_c,
\ee
with $\zeta'=(g'_{\mn},\chi')$ a collective variable for the fluctuating metric $g'_{\mn}$ and scalar field $\chi'$. Our starting point is the defining equation for the Green's function
\be\label{I6}
{\cal D}G(x,y)=E(x,y).
\ee
The differential operator $\D$ acts on $x$, and $E$ is the unit matrix in field space which may contain appropriate projectors in case of constrained fields. The relation to the quantum effective action is given by $\D=\Gamma^{(2)}$, with $\Gamma^{(2)}$ the second functional derivative of the effective action $\Gamma$. The propagator equation \eqref{I6} expresses directly the exact basic identity for the quantum effective action $\Gamma^{(2)}G=1$. 

The propagator equation \eqref{I6} is an inhomogeneous differential equation due to the unit matrix on the r.h.s.. This restricts the normalization of $G$ and plays precisely the role of the normalized commutator relations in the usual treatment of quantum field operators in the vacuum state. We emphasize, however, that eq. \eqref{I6} only involves ``classical'' variables $G(x,y)$ - no operators appear in this formulation. The identity $\Gamma^{(2)}G=1$ relates the fluctuation properties directly to the second functional derivative of $\Gamma$. Non-linear field transformations between different frames can therefore be performed in a straightforward manner.

The present paper is organized as follows. In sect. \ref{Quantum effective action and field equations} we display the effective action for variable gravity and the field equations derived from it. For homogeneous and isotropic cosmology we express the field equations in terms of frame invariant variables. Solutions apply to all frames that are related by conformal rescalings of the metric. They are easily translated into each individual frame. In sect. \ref{Correlation function} we formulate the propagator equation \eqref{I6} for variable gravity. For a homogeneous and isotropic background the mode expansion of fluctuations is again formulated in a frame invariant way. 

In sect. \ref{Graviton correlation}  we turn to the correlation function for the graviton. The relevant propagator equation is expressed in terms of frame invariant variables. We discuss the general solution and concentrate on initial values corresponding to the Bunch-Davies vacuum. We discuss the time evolution of the power spectrum, establish the ``freeze condition'' $k\ll\hat \h$, and discuss the exact solution for geometries $\p\hat \h=(1+\nu)\hat \h^2$ with constant $\nu$. The exact tensor spectral index is found as $n_T=2\nu/(1+\nu)$. Approximate scale invariance of the spectrum obtains for $|\nu|\ll 1$ and can be realized even for geometries without a geometrical horizon, $H^{-1}\to\infty$. 

Sect. \ref{Field relativity} turns to a brief discussion of the relevant aspects of field relativity and maps our results to the Einstein frame with fixed Planck mass. In sect. \ref{Vector and scalar correlations} we address the vector fluctuations of the metric. Even though there is no propagating gauge invariant vector fluctuation, the correlation function in the vector part does not vanish. The gauge invariant vector correlation is proportional $k^{-2}\delta(\eta-\eta')$. We briefly discuss general aspects of scalar fluctuations without entering detailed computations. Using field relativity the scalar fluctuations in variable gravity can be obtained from the ones in the Einstein frame.

In sect. \ref{Primordial fluctuations from flat space} we discuss an example of variable gravity where a realistic fluctuation spectrum is generated for a geometry of flat Minkowski space. This demonstrates clearly that no geometric horizon is needed for a realistic fluctuation spectrum. Our conclusions are presented in sect. \ref{Conclusions}. Several computational aspects are displayed in appendices.

\section{Quantum effective action and field equations}
\label{Quantum effective action and field equations}

In this section we discuss the quantum effective action and the field equations for variable gravity. For homogeneous and isotropic cosmology we write the field equations in a form that is invariant under conformal scalings of the metric.

\subsection{Effective action and field relativity}

We consider here the most general effective action $\Gamma$ for a scalar field $\chi$ and the metric $g_{\mu\nu}$ that contains at most two derivatives. For this model of ``variable gravity'' \cite{CWVG} one has
\be\label{166}
\Gamma=\int_x\sqrt{g}
\left\{-\frac12F(\chi)R+\frac12 K(\chi)\partial^\mu\chi\partial_\mu\chi+V(\chi)\right\}.
\ee
(For the simplified discussion in the introduction we have used $F=\chi^2$.) Physical observables can be expressed by $\Gamma$ and its functional derivatives. The first derivative determines the expectation values of the metric and the scalar field as solutions of the field equations. The second derivative constitutes the inverse propagator, such that the two-point correlation function or power spectrum can be obtained by its inversion \cite{CWneu,CW2}. Higher derivatives are directly related to higher order correlation functions as the bispectrum. 

Let us consider non-linear transformations of the metric $g_{\mu\nu}$ and the scalar field $\chi$. Physical observables are given as functionals of fields or correlation functions. Their transformation properties under a change of field variables follow directly by employing the transformed fields in the corresponding functionals. Physical observables cannot depend on the choice of fields used to describe them - a property called ``field relativity'' in ref. \cite{CWVG,VG30}. All effective actions that can be obtained from each other by field transformations are equivalent if observables are transformed appropriately.

We concentrate here on field-dependent conformal transformations of the metric,
\be\label{F2-a}
g_{\mn}=w^2(\chi)g'_{\mn}.
\ee
The curvature scalar transforms as
\be\label{F2-b}
R=w^{-2}
\big[R'-6g'^{\mn}
(D_\mu D_\nu\ln w+\partial_\mu\ln w\partial_\nu\ln w)\big],
\ee
where $D_\mu$ denotes a covariant derivative involving the connection of the metric $g'_{\mn}$. The effective action \eqref{166} maintains its form, with 
\be\label{F2-c}
F'=w^2F~,~V'=w^4V,
\ee
and 
\ba\label{F2-d}
K'&=&w^2
\left[K-6F\frac{\partial\ln w}{\partial\chi}
\left(\frac{\partial\ln w}{\partial\chi}+\frac{\partial\ln F}{\partial\chi}\right)\right]\nn\\
&=&\frac{F'}{F}K+\frac32 F'
\left[\left(\frac{\partial\ln F}{\partial\chi}\right)^2-
\left(\frac{\partial\ln F'}{\partial\chi}\right)^2\right].
\ea
Variable gravity models where $(F,V,K)$ and $(F',V',K')$ can be related by the relations \eqref{F2-c}, \eqref{F2-d} are equivalent. They correspond to different ``frames'' that describe the same physical situation. Nevertheless, the geometrical picture may look very different in different frames. 

The combinations
\be\label{11A}
\hat V=\frac{V}{F^2},
\ee
and
\be\label{Z3}
\hat K=\frac{K}{F}
+\frac{3}{2F^2}\left(\frac{\partial F}{\partial\chi}\right)^2,
\ee
are invariant under the conformal transformations \eqref{F2-a}. The physical content of a given model is specified by the two invariant functions $\hat V(\chi)$ and $\hat K(\chi)$, while different frames or pictures involve in addition the specification of $F(\chi)$. (Various quantities that are invariant under conformal transformations have been discussed earlier, see refs. \cite{CWEU,JK} for recent discussions.)

\subsection{Field equations}

In the absence of additional matter and radiation the gravitational field equations for variable gravity read \cite{CWVG}
\ba\label{292}
&&F(R_{\mn}-\frac12 Rg_{\mn})+D^2 Fg_{\mn}-D_\mu D_\nu F\nn\\
&&+\frac12K\partial^\rho\chi\partial_\rho\chi g_{\mn}-K\partial_\mu\chi\partial_\nu\chi+ Vg_{\mn}=0,
\ea
while the scalar field equation is given by $(D^2=D_\mu D^\mu)$
\be\label{X2-a}
KD^2\chi+\frac12\frac{\partial K}{\partial\chi}\partial^\mu\chi\partial_\mu \chi=\frac{\partial V}{\partial\chi}-\frac12\frac{\partial F}{\partial \chi}R.
\ee

For ``background cosmology'' we concentrate on a homogeneous and isotropic geometry with metric
\be\label{C2-a}
\bar g_{\mn}=a^2(\eta)\eta_{\mn},
\ee
where the scale factor $a(\eta)$ is dependent on conformal time $\eta$. The scalar field $\bar\chi(\eta)$ is homogeneous as well.  For a homogeneous and isotropic cosmology the two independent gravitational field equations are given by  
\ba\label{172}
F\partial_\eta\h &=&\frac{a^2}{3}V-\frac{K}{3}(\partial_\eta\bar \chi)^2-\frac12\partial^2_\eta F,\nn\\
F\h^2&=&\frac{a^2}{3}V+\frac K6(\partial_\eta\bar \chi)^2-\h\partial_\eta F,
\ea
with conformal Hubble parameter
\be\label{172A}
\h=\p\ln a=Ha.
\ee
The scalar field equation becomes
\ba\label{Y2-a}
&&K(\p^2+2\h\p)\chi+\frac12\frac{\partial K}{\partial\chi}(\p\chi)^2\nn\\
&&\qquad~ +a^2\frac{\partial V}{\partial \chi}-3\frac{\partial F}{\partial\chi}
(\h^2+\p\h)=0.
\ea

For the metric \eqref{C2-a} the conformal scaling \eqref{F2-a} results in a multiplicative rescaling of the scale factor
\be\label{I2-a}
a'=\frac{a}{w}=\sqrt{\frac{F}{F'}}a.
\ee
We can therefore form the combination
\be\label{I2-b}
A=\sqrt{F}a
\ee
which is invariant under conformal transformations. It measures the scale factor in units of the variable Planck length.
The invariant expansion parameter
\be\label{Z1}
\hat \h=\partial_\eta\ln A.
\ee
describes the change of the dimensionless ratio between the scale factor and the variable Planck length,
\be\label{170}
\hat \h=\frac12\partial_\eta\ln (a^2F)=\h+\frac12\partial_\eta\ln F. 
\ee

The two gravitational field equations can be written in an explicitly frame invariant form
\be\label{M2-a1}
2\hat\h^2+\p\hat \h=A^2\hat V,
\ee
and 
\be\label{M2-b}
\hat \h^2-\p\hat \h=\frac{\hat K}{2}
(\p\bar \chi)^2.
\ee
The same holds for the scalar field equation 
\be\label{M2-c}
\hat K(\p^2+2\hat \h\p)\chi+\frac12\frac{\partial \hat K}{\partial \chi}(\p\chi)^2+A^2
\frac{\partial\hat V}{\partial\chi}=0.
\ee
Eqs. \eqref{M2-a1} - \eqref{M2-c} are a coupled system of differential equations for the two functions $A(\eta)$ and $\bar\chi(\eta)$. As usual, only two of them are independent. We may actually insert the homogeneous isotropic metric and scalar field directly into the effective action. One finds
\be\label{M2-d}
\Gamma=i\int_x
\left\{3(\p A)^2-\frac{A^2\hat K}{2}(\p\chi)^2+A^4\hat V\right\},
\ee
and we can obtain the two independent field equations by variation with respect to $\chi(\eta)$ and $A(\eta)$. 

The solution for $A(\eta)$ and $\bar\chi(\eta)$ holds for arbitrary frames. For a given frame (given choice of $(F,V,K)$) the scale factor $a(\eta)$ can then be extracted from eq. \eqref{I2-b}. The formulation in terms of the conformally invariant scale factor $A(\eta)$ permits also an easy map of cosmological solutions from one frame to another. Since $\bar\chi(\eta)$ is the same in all frames, we can compute $F(\eta)$ and then $A(\eta)$ for a given solution $a(\eta)$ by employing eq. \eqref{I2-b}. Using $F'(\eta)$ in the new frame one extracts $a'(\eta)$ from eq. \eqref{I2-a}. In particular, one may choose the Einstein frame where 
\be\label{C3-a}
F=M^2~,~\hat K=\frac{K}{M^2}~,~\hat V=\frac{V}{M^4},
\ee
with $M$ the fixed reduced Planck mass. Cosmological solutions in the Einstein frame can be translated to the corresponding solution in any other frame.

\section{Correlation function}
\label{Correlation function}

We next turn to the computation of the correlation function which encodes the information about the power spectrum of the primordial cosmic fluctuations. The correlation function is the inverse of the second functional derivative of $\Gamma$. This allows a discussion without invoking quantum vacua - knowledge or assumptions about the form of $\Gamma$ are sufficient. In particular, the effective action \eqref{166} for variable gravity specifies uniquely a time evolution equation for the correlation function. Particular solutions are selected by initial conditions. The power spectrum equals the correlation function for equal time arguments. 

\subsection{Propagator equation}

The basic identity for the computation of the correlation function $G(x,y)$ is the relation
\be\label{PA}
\Gamma^{(2)}G=1,
\ee
or
\be\label{PB}
\int_y\Gamma^{(2)}(x.y)G(y,z)=E(x,z).
\ee
(For a simple unconstrained scalar field one has $E(x,y)=\delta(x-y).$) The second functional derivative $\Gamma^{(2)}$ is matrix valued, e.g.
\be\label{PC}
\Gamma^{(2)\mn\rt}_{gg}(x,y)=
\frac{\delta^2\Gamma}{\delta g_{\mn}(x)\delta g_{\rt}(y)},
\ee
and similar for the scalar and mixed components. The second functional derivative for the action \eqref{166} of variable gravity is computed in appendix A. Also the correlation function $G$ is matrix valued, and the unit matrix on the r.h.s. of eq. \eqref{PB} may contain appropriate projectors. More details of the formal setting can be found in appendix B. 

Expanding the metric and scalar field 
\ba\label{PD}
g_{\mn}(x)&=&\bar g_{\mn}(\eta)+h_{\mn}(\eta,\vec x),\nn\\
\chi(x)&=&\bar \chi(\eta)+\delta\chi(\eta,\vec x),
\ea
the connected two point function for the metric reads
\be\label{PE}
G^{gg}_{\mn\rt}(x,y)=\kl h'_{\mn}(x)h'_{\rt}(y)\kr_c,
\ee
and similar for the other components involving $\delta\chi$. (Here the fluctuating metric in the functional integral is expanded similar to eq. \eqref{PD}, $g'_{\mn}=\bar g_{\mn}+h'_{\mn}$, and we identify the background metric with the argument of the effective action, $\kl g'_{\mn}\kr=\bar g_{\mn}$. If $\bar g_{\mn}$ and $\bar\chi$ do not obey the field equations \eqref{172}, \eqref{Y2-a}, the source terms in the formulation of the functional integral do not vanish.) For the computation of the second functional derivative $\Gamma^{(2)}$ we evaluate the effective action \eqref{166} in second order in the fluctuations $h_{\mn}$ and $\delta\chi$. The resulting detailed expression for $\Gamma_2$ is displayed in appendix A.

\subsection{Mode expansion}

It is convenient to work in three-dimensional momentum space
\be\label{H2}
h_{\mu\nu}(x)=h_{\mu\nu}(\eta,\vec x)=\int \frac{d^3 k}{(2\pi)^3}
e^{i\vec k\vec x}h_{\mu\nu}(\eta,\vec k).
\ee
With respect to the $SO(3)$-symmetry of rotations in space we decompose the momentum modes of the metric into the graviton $\gamma_{mn}$, two vectors $V_m$ and $W_m$, and four scalars $A,B,C,D$, according to 
\ba\label{PF}
h_{00}&=&-2Aa^2~,~h_{m0}=a^2(W_m+ik_mD),\\
h_{mn}&=&a^2(\gamma_{mn}+ik_mV_n+ik_nV_m-2k_mk_n B+2\delta_{mn}C),\nn
\ea
with 
\ba\label{PG}
k^mV_m&=&0~,~k^mW_m=0,\nn\\
k^m\gamma_{mn}&=&0~,~\delta^{mn}\gamma_{mn}=0.
\ea
The conformal transformation \eqref{F2-a} is already accounted for by the rescaling \eqref{I2-a} of the scale factor, such that the fields $\gamma_{mn},W_m,V_n,A,B,C$ and $D$ are invariant under frame transformations. Since scalars, vectors and the traceless divergence free tensor do not mix in quadratic order we can solve the propagator equation \eqref{PB} separately for these components.

A general coordinate transformation changes the background metric $\bar g_{\mn}$. Part of the metric fluctuations $h_{\mn}$ correspond to fields that can be generated from $\bar g_{\mn}$ by an infinitesimal coordinate transformation. The remaining fields that do not correspond to these ``gauge directions'' are the ``gauge invariant'' scalar Bardeen potentials \cite{Bar}
\ba\label{PH}
\Phi&=&C-\h(\p B-D),\nn\\
\Psi&=&A-(\p+\h)(\p B-D),
\ea
the vector 
\be\label{PI}
\Omega_m=W_m-\p V_m,
\ee
and the graviton $\gamma_{mn}$. A similar ``gauge invariant'' scalar fluctuation is defined as
\be\label{PJ}
X=\delta\chi-\p\bar \chi(\p B-D).
\ee
While $\Omega_m,X$ and $\Phi-\Psi$ are invariant under conformal transformations, one has
\ba\label{PK}
\Phi'+\Psi'&=&\Phi+\Psi-2(\h'-\h)(\p B-D)\nn\\
&=&\Phi+\Psi+2\p \ln w(\p B-D).
\ea

\section{Graviton correlation}
\label{Graviton correlation}

The tensor modes of the cosmic fluctuation spectrum are extracted from the correlation function of the graviton $\gamma_{mn}$.

\subsection{Evolution equation for graviton fluctuations}

On the quadratic level the graviton $\gamma_{mn}$ does not mix with the scalar and vector fluctuations. For the computation of the graviton propagator we can therefore employ the metric 
\be\label{138A}
h_{mn}=a^2\gamma_{mn}~,~h_{m0}=0~,~h_{00}=0.
\ee
This implies $h=0,~h^\nu_{\mu;\nu}=0$. Inserting the ansatz \eqref{138A} into the effective action and expanding to second order in $\gamma_{mn}$ we find in appendix C
\be\label{J1}
\Gamma^{(\gamma)}_2=\frac12\int_{\eta,k}\gamma^*_{mn}(\eta,k)\Gamma^{(2)mnpq}_\gamma(k)\gamma_{pq}(\eta,k),
\ee
with 
\be\label{J2}
\Gamma^{(2)mnpq}_\gamma(k)=\frac i4
P^{(\gamma)mnpq}
\big\{A^2(\p^2+2\hat \h\p+k^2)+\Delta_\gamma\big\},
\ee
and
\be\label{J3}
\Delta_\gamma=2A^2(\hat \h^2+2\p\hat \h)-2A^4\hat V+A^2\hat K(\p\bar \chi)^2.
\ee
The projector $P^{(\gamma)}$ depends on $\vec k$ and is given by 
\be\label{159A}
P^{(\gamma)}_{mnpq}=\frac12(Q_{mp}Q_{nq}+Q_{mq}Q_{np}-Q_{mn}Q_{pq}),
\ee
with $Q_{mn}$ given by
\be\label{E3B}
Q_{mn}=\delta_{mn}-\frac{k_mk_n}{k^2}.
\ee
(For the projectors $Q$ and $P^{(\gamma)}$ we raise indices exceptionally with $\delta^{mn}$.)

We observe that only the frame invariant quantities $A,\hat \h,\hat V, \hat K$ appear in eq. \eqref{J2}. We can therefore employ for all frames the same propagator equation for the graviton,
\be\label{Z2}
\frac{iA^2}{4}(\partial^2_\eta+2\hat \h\partial_\eta+k^2+\Delta_\gamma)
G^{\gamma\gamma}_{mnpq}=P^{(\gamma)}_{mnpq}\delta(\eta-\eta').
\ee
If the background metric and scalar field obey the field equations \eqref{M2-a1}, \eqref{M2-b} the term $\Delta_\gamma$ vanishes. We assume this in the following. Rotation symmetry implies for a traceless and divergence free symmetric tensor $(k=|\vec k|)$
\ba\label{S8}
G^{\gamma\gamma}_{mnpq}(\vec k,\eta,\eta')=P_{mnpq}^{(\gamma)} G_{\rm grav}(k,\eta,\eta'),
\ea
such that the propagator equation becomes
\be\label{S9}
(\partial^2_\eta+2\hat {\cal H}\partial_\eta+k^2)G_{grav}(k,\eta,\eta')=-
\frac{4i}{A^2}\delta(\eta-\eta').
\ee
Its solution determines $G^{\gamma\gamma}$ for all frames. The graviton contribution to the metric correlation $G_{\mn\rt}$ obtains from $G^{\gamma\gamma}_{mnpq}$ by multiplication with $a^2(\eta)a^2(\eta')$ and depends on the frame. (It vanishes for all components where at least one index is zero.)

We recall that the (partial) normalization of the correlation function arises from the inhomogeneous term on the r.h.s. of eq. \eqref{S9}. This replaces the commutator relations in the approach based on quantum fields in vacuum. Still, eq. \eqref{S9} is a differential equation whose solution depends on initial values. We may set these initial values for large negative $\eta$ or $\eta\to-\infty$. 

\subsection{General solution for graviton correlation}

The general solution of eq. \eqref{S9} has been discussed extensively in ref. \cite{CWneu}, \cite{CW2}. It is the same as for a massless scalar field in a fixed background geometry. For $\eta>\eta'$ the solution reads
\ba\label{S10}
&&G_{grav}(k,\eta,\eta')={2\big(\alpha(k)+1\big)}
w^-_k(\eta)w^+_k(\eta')\nn\\
&&\quad +{2\big(\alpha(k)-1\big)}w^+_k(\eta)w^-_k(\eta')\\
&&+\quad {4\zeta(k)}w^+_k(\eta)w^+_k(\eta')+
{4\zeta^*(k)}w^-_k(\eta)w^-_k(\eta'),\nn
\ea
with mode functions obeying
\be\label{WA}
w^+_k(\eta)=\big (w^-_k(\eta)\big)^*,
\ee
and normalization such that for $k\eta\to-\infty$ one has
\be\label{WB}
\lim_{k\eta\to-\infty}w^-_k(\eta)=
\frac{1}{A(\eta)\sqrt{2k}}e^{-ik\eta}.
\ee
For each $k$-mode one has three real integration constants, corresponding to real $\alpha(k)$ and complex $\zeta(k)$. They are not determined by the partial normalization due to the r.h.s. of eq. \eqref{S9}. The general solution of the propagator equation remains an initial value problem. The solution for $\eta<\eta'$ obtains from eq. \eqref{S10} by interchanging $\eta\leftrightarrow\eta'$

For Bunch-Davies initial conditions \cite{BD}, which correspond to the scaling correlation of ref. \cite{CW2}, one has $\alpha(k)=1,\zeta(k)=0$, such that 
\be\label{134}
G\gr (k,\eta,\eta')={4}w^-_k(\eta)w^+_k(\eta').
\ee
In the limit $k\eta,k\eta'\to-\infty$ the graviton correlation becomes then
\be\label{165A}
\lim_{k\eta\to-\infty} G_{\rm grav}(k,\eta,\eta')=\frac{2}{kA(\eta)A(\eta')}
e^{-ik(\eta-\eta')}.
\ee
For $A(\eta)=A(\eta')=M$ this coincides with the flat space correlation. Thus the ``Bunch-Davies vacuum'' corresponds to initial conditions for which the correlation function is given by the one for flat space. 

We will in the following concentrate on this initial condition. This is motivated by the following observations: The observable modes obey in the early stages of their evolution the relation $k^2\gg\hat \h^2$. Neglecting the term $\sim\hat\h\p$ in eq. \eqref{S9}, and taking correspondingly a value of $A$ which is constant on the relevant time scale, eq. \eqref{S9} reduces to the free field equation in Minkowski space. The Bunch-Davies initial condition \eqref{165A} corresponds to the Lorentz-invariant propagator in flat space. While the approximation \eqref{166} for the quantum effective action provides for no mechanism for an ``equilibration'' of an arbitrary initial propagator towards the Lorentz-invariant one \cite{CWneu,CW2}, such a mechanism may be found by extending $\Gamma$ beyond the approximation \eqref{166}. Such extensions are indeed expected from fluctuation effects in interacting quantum field theories, cf. refs. \cite{Wein,Sha,TW,BV,ProS,One1,One2,Boy,Akh,Poly,Bur,GS,KH,GGG}. The issue is similar to the understanding of thermalization of arbitrary initial correlation functions in flat space, for which powerful functional methods have been developed \cite{CWET1,CWET2,BW,ABW1,ABW2,BC,AB,BR,AB2,PT,BBWPT}. Equilibration of the short distance tail of the propagator to the Lorentz-invariant one is sufficient to guarantee the ``Bunch-Davies-vacuum'' for all later times \cite{CW2}. This scenario would indeed realize eq. \eqref{165A} for the observable modes, provided inflation lasts long enough before the freezing of the observable modes.

\subsection{Primordial tensor modes}

The tensor power spectrum is defined by the equal time correlation function
\be\label{136}
\Delta^2_T(k,\eta)=\frac{k^3}{\pi^2}G\gr(k,\eta,\eta).
\ee
Correspondingly, the tensor spectral index obeys
\be\label{137}
n_T=\frac{\partial \ln \Delta^2_T}{\partial\ln k}.
\ee

We employ eq.  \eqref{134}, with mode functions obeying 
\be\label{150}
(\partial^2_\eta+2\hat \h\partial_\eta+k^2)w^\pm_k(\eta)=0.
\ee
It is instructive to consider
\be\label{151a}
v^\pm_k(\eta)=A(\eta)w^\pm_k(\eta)
\ee
with evolution equation
\be\label{152}
(\partial^2_\eta-\frac{\partial^2_\eta A}{A}+k^2)v^\pm_k(\eta)=0.
\ee

We distinguish two regimes: For $k\gg\hat \h$ the term $\sim \partial^2_\eta A/A$ is negligible and the solution is oscillating
\be\label{153}
w^-_k(\eta)=\frac{c_k}{A(\eta)}e^{-ik\eta}.
\ee
For $k\ll \hat\h$ the term $|\p^2 A/A|$ dominates over $k^2$, and the modes evolve according to 
\be\label{154}
(\partial^2_\eta+2\hat \h \partial_\eta)w^\pm_k=0.
\ee
The solutions approach constants, $w^\pm_k=$const. This corresponds to the freezing of fluctuations discussed in the introduction. 

Typical graviton modes start with $k\gg\hat\h$. The mode functions are then given by eq. \eqref{153}, with $c_k=1/\sqrt{2k}$. In primordial cosmology $\hat \h$ increases. Once a given mode reaches the regime $k\ll\hat\h$ the mode functions (and the associated power spectrum) freeze according to eq. \eqref{154} or
\be\label{155}
\partial_\eta w^-_k=\frac{d_k}{A^2(\eta)}.
\ee
In this regime the modes keep a fixed amplitude, corresponding to the value at the time when $\hat \h\approx k$. Afterwards $\hat \h$ decreases again in later stages of cosmology, corresponding to the evolution after the end of inflation in the usual picture. Once a mode enters again the regime $k\gg\hat \h$ at a time $\eta_{hc}$, it starts again the damped oscillation \eqref{153}, now with $c_k$ given by the value at $\eta_{hc}$, which is almost the same as the one when the mode has first reached the regime $k\ll\hat \h$. Due to the subsequent decrease of the tensor fluctuation spectrum $\sim A^{-2}(\eta)$, only the modes that enter again the regime $k\gg\hat \h$ near the present epoch may be observable in practice. 

We emphasize that the evolution of the graviton fluctuations - oscillation, freeze and oscillation again - is entirely characterized in terms of the variable
\be\label{55A}
\hat k=\frac{k}{\hat \h}.
\ee
This is the appropriate dimensionless quantity for the description of the dynamics of a given fluctuation. A freeze of the fluctuation with preservation of memory happens whenever $\hat k\ll 1$. This does not need the presence of a geometric horizon - $H^{-1}$ may be arbitrarily large if $\p\ln F$ is large enough as compared to $k$. Our discussion is manifestly frame invariant and yields the same power spectrum in all frames related by the conformal scaling \eqref{F2-a}. 

\subsection{Cosmologies with constant tensor spectral index}

It is instructive to consider geometries obeying
\be\label{156}
\partial_\eta\hat \h =(1+\nu)\hat\h^2.
\ee
With $\partial^2_\eta A/A=\partial_\eta\hat\h+\hat \h^2$ the mode equation \eqref{152} reads
\be\label{157}
[\partial^2_\eta+k^2-(2+\nu)\hat\h^2]v^-_k=0.
\ee
For constant $\nu$ the solution is given \cite{CW2} by
\be\label{158}
v^-_k=\frac{1}{\sqrt{2k}}e^{-ik\eta}\left[1+\frac{i}{y}(1+\nu\tilde f(iy)\right]^{\frac{1}{1+\nu}},
\ee
with
\be\label{159}
y=\hat k=\frac {k}{\hat\h}
\ee
and $\tilde f(x)$ a smooth function (denoted by $(f-1)/\nu$ in ref. \cite{CW2}) varying between $\tilde f(0)=0.273$ and $\tilde f(x\to \infty)=0.5$.

The regime $k\ll\hat \h$ corresponds to $y\to0$ or
\be\label{WC}
w^-_k=\frac{\hat C}{\sqrt{2}}(i+0.273 i\nu)^{\frac{1}{1+\nu}}
k^{-\frac{3+\nu}{2(1+\nu)}}
e^{-ik\eta}.
\ee
Here $\hat C$ is constant for geometries with constant $\nu$
\be\label{WD}
\hat C=A^{-1}\hat \h^{\frac{1}{1+\nu}}~,~\p\hat C=0.
\ee
One infers for the tensor  power spectrum
\ba\label{162}
\Delta^2_T&=&\frac{2\hat C^2}{\pi^2}|b_0|^2
k^{\frac{2\nu}{1+\nu}},\nn\\
|b_0|^2&=&(1+0.273\nu)^{\frac{2}{1+\nu}}\approx 1+0.55\nu.
\ea
The tensor spectral index is given by the exact result
\be\label{163}
n_T=\frac{2\nu}{1+\nu}.
\ee
Since $\hat C$ is constant we can evaluate $\Delta^2_T$  at the time when $\hat \h=k$, namely
\be\label{WE}
\Delta^2_T=\frac{2|b_0|^2k^2}{\pi^2A^2_c},
\ee
with $A_c$ the value of $A$ at the time when $\hat \h=k$. 

We repeat that $\hat \h$ may increase purely due to an increase of the logarithmic growth rate for the Planck mass $\sqrt{F}$, even for flat space with a constant scale factor $a$. For the particular case of flat space, $\h=0$, one has
\be\label{WF}
\nu=\frac{2F\p^2 F}{(\p F)^2}-3.
\ee
An almost scale invariant spectrum with $|\nu|\ll1$ does not need an expanding universe. Indeed, for constant $a$ the scaling $F\sim-\eta^{-2}$ results in $\nu_T=0$. (A similar observation has been made in ref. \cite{Pi,ML1,ML2}.) This amounts to $\hat \h=-1/\eta$. In the Einstein frame such a solution will correspond to constant $H$.

\section{Field relativity}
\label{Field relativity}

Since eq. \eqref{PA} expresses the evolution equation for the correlation functions directly in terms of the quantum effective action we can make arbitrary non-linear field transformations in $\Gamma$. Physical observables do not depend on the choice of fields - a property called ``field relativity'' in ref. \cite{VG30,CWVG}. We can compute quantities as the tensor spectral index $n_T$ for an arbitrary choice of fields or arbitrary ``frames''. The physical equivalence of different frames can be established \cite{VG3,VG4,VG117,VG101,VG118,VG119,VG115,VG116,CWEU} directly if observables are expressed in terms of the quantum effective action and its functional derivatives (field equations, correlation functions). The equivalence of the evolution of mode functions can be based on the linearized field equations if the background solution is itself a solution of the field equations. Beyond this, our approach also allows for a frame invariant discussion of the evolution of mode functions if the background does not obey the field equations (presence of sources). In this case a term $\Delta_\gamma w^\pm_k$ has to be added to the left hand side of eq. \eqref{150}. 

The direct translation of properties of a ``quantum vacuum'' between different frames can be cumbersome. We bypass this issue by basing the whole discussion on the inversion of the second functional derivative of $\Gamma$. On this level field transformations are straightforward. The partial normalization of the correlation function is now ensured by the inhomogeneous propagator equation \eqref{S9}, and the selection of the vacuum that is analogue to the Bunch-Davies vacuum is operated by the initial condition \eqref{165A}. Since our discussion of the propagator equation and its solution only involves frame-invariant quantities, the equivalence of different frames is manifest and the mapping between them becomes very simple. 

In particular, a Weyl scaling \cite{VG123} of the metric
\be\label{Z5}
g'_{\mu\nu}=\frac{F(\chi)}{M^2}g_{\mu\nu}
\ee
preserves the general form \eqref{166} of the effective action for variable gravity, while bringing the coefficient in front of the curvature scalar to a constant $F'=M^2$. (This is called ``Einstein frame''.) In the Einstein frame the rescaled potential or kinetial read \cite{CWVG}
\be\label{Z6}
V'=\frac{M^4V}{F^2}~,~K'=M^2
\left\{\frac KF+\frac32
\left(\frac{\partial\ln F}{\partial\chi}\right)^2\right\}.
\ee
While $\chi$ and $\eta$ are not affected by the Weyl scaling, the metric transformation is realized by $a(\eta)\to a'(\eta)$
\be\label{Z7}
a'(\eta)=\frac{\sqrt{F(\chi)}}{M}a(\eta).
\ee
The coordinates $\vec x$ and the comoving wave vector $\vec k$ do not change either. We conclude that the primordial fluctuation spectrum can indeed be computed in any arbitrary frame, and in particular in the Einstein frame, as advocated since a long time \cite{VG71}. 

In the Einstein frame one has $A=Ma,\hat \h=\h=Ha$. The change between the regimes with $k\gg\hat \h$ and $k\ll\hat \h$ corresponds to horizon crossing, e.g. $\hat \h=k$ is reached for $k/a=H$. With $A_c=Ma_c=Mk/H_{hc}$, where $H_{hc}$ is the value of $H$ at horizon crossing, $H_{hc}=H(y=1)$, eq. \eqref{WE} yields the well known relation between the tensor amplitude and the Hubble parameter at horizon crossing
\be\label{C4-a}
\Delta^2_T=\frac{2|b_0|^2H^2_{hc}}{\pi^2M^2}.
\ee
(The factor $|b_0|^2$ is very close to one and usually omitted.) Our discussion of different frames has revealed, however, that there is no profound relation with the existence of a geometric horizon. The same physical situation arises in frames where no horizon is present for the geometry.

\section{Vector and scalar correlations}
\label{Vector and scalar correlations}

For a constant Planck mass it is well known that the solution of the linearized field equation does not allow for a propagating gauge invariant vector fluctuation, e.g. they imply
\be\label{69A}
\Omega_m=0.
\ee
We extend this to variable gravity. Nevertheless, the metric correlation function in the vector channel does not vanish. This simply follows from the inhomogeneous term on the r.h.s. of eq. \eqref{PB} which does not vanish once projected on the vector modes. The situation is similar in the scalar sector. Out of the three gauge invariant scalar modes $X,\Phi$ and $\Psi$ only one is propagation. The correlation function cannot be reduced to the propagating mode, however.

\subsection{Vector correlation}

The vector fluctuations do not mix with the graviton or the scalar fluctuations. The relevant inverse propagator obtains by inserting the ansatz
\ba\label{Z8}
h_{mn}&=&ia^2(k_mV_n+k_nV_m),\nn\\
h_{0m}&=&a^2W_m~,~h_{00}=0,
\ea
into the expansion \eqref{C1} of the effective action for variable gravity. The quadratic effective action is computed in appendix D for arbitrary vector fluctuations. In terms of $\Omega_m=W_m-\partial_\eta V_m$ one finds
\be\label{A20-1}
\Gamma^{(V)}_2=-\frac i4\int_{\eta,k}
A^2k^2\Omega^*_m Q^{mn}\Omega_n+\Delta\Gamma^{(V)}_2,
\ee
with 
\ba\label{A20-2}
\Delta\Gamma^{(V)}_2&=&-\frac{i}{4}\int_{\eta,k}A^2Q^{mn}
\big\{[2A^2\hat V-4\partial_\eta\hat \h-2\hat \h^2\nn\\
&&-\hat K(\partial_\eta\bar \chi)^2](W^*_mW_n+k^2V^*_mV_n)\nn\\
&&+4[\p\hat\h+2\hat\h^2-A^2\hat V]W^*_mW_n\big\}.
\ea
It involves only the frame invariant quantities $A^2,\hat V,\hat K,\bar\chi$ and the frame-invariant fluctuations $W_m$ and $V_m$. It therefore takes the same form in all frames that are related by a conformal rescaling of the metric. 

If the background obeys the field equations the term $\Delta\Gamma^{(V)}_2$ vanishes. We will concentrate on this case in the following. Then $\Gamma_2$ depends only on the gauge invariant vector field $\Omega_m$. The linearized field equation in the vector channel reads
\be\label{72A}
A^2k^2Q^{mn}\Omega_n=0,
\ee
implying eq. \eqref{69A}. The second functional derivative with respect to $\Omega_m$ becomes
\be\label{A5-a}
\Gamma^{(2)mn}_\Omega=-\frac i2A^2k^2Q^{mn}.
\ee
It does not contain any derivatives with respect to $\eta$. For the vector correlation
\be\label{A5-b}
\kl \Omega'_m(\eta,\vec k)\Omega'^{*}_n(\eta',\vec k')\kr=
G^{\Omega\Omega}_{mn}(\vec k,\eta,\eta')\delta(k-k')
\ee
rotation symmetry requires
\be\label{A5-c}
G^{\Omega\Omega}_{mn}(\vec k,\eta,\eta')=G_\Omega(k,\eta,\eta')Q_{mn}. 
\ee
The propagator equation reads
\be\label{A5-d}
-\frac i2 A^2k^2G_\Omega(k,\eta,\eta')=\delta(\eta-\eta').
\ee

The solution of the propagator equation is not expressed by mode functions, but simply reads
\be\label{A5-e}
G_\Omega(k,\eta,\eta')=\frac{2i}{A^2k^2}\delta(\eta-\eta').
\ee
While the vector correlation vanishes for all $\eta\neq\eta'$, it differs from zero for $\eta=\eta'$. The correlation function \eqref{A5-e} implies an instantaneous response of the metric to vector perturbations. This shows in the language of correlation functions that one has no propagating vector field, but rather a type of auxiliary field. The momentum behavior is $k^{-2}$, corresponding to $r^{-1}$ in position space. This is similar to Newton's potential. 

\subsection{Scalar correlation}

The scalar sector is technically the most cumbersome due to the presence of several scalar modes. In the view of field relativity one may first solve the fluctuation problem in the Einstein frame and then map the solution to other frames. For the Einstein frame the mode equation for the propagating scalar fluctuation has been discussed extensively in the literature. For backgrounds solving the field equations the propagator equation for the propagating scalar mode and its general solution have been established in ref. \cite{CW2}. For these reasons we omit in the present paper a detailed treatment of the scalar fluctuations. A computation of the propagator equation in terms of frame invariant variables, similar to the graviton and vector fluctuations, would complete the explicitly frame-invariant setting of the fluctuation problem. The correlation function for the non-propagating scalar modes is similar to eq. \eqref{A5-e} and describes Newton's potential.

\section{Primordial fluctuations from flat space}
\label{Primordial fluctuations from flat space}

As a specific example we discuss primordial fluctuations in a geometry that approaches flat Minkowski space. We take the crossover model for variable gravity of ref. \cite{CWCG}, and discuss early cosmology in the ``primordial flat frame''. For this model one has for the relevant range of small $\chi$
\ba\label{C6a}
F&=&\chi^2~,~V=\bar\lambda \chi^4\ln \frac{\bar m}{\chi},\nn\\
K&=&\frac{2}{\ln \left(\frac{\bar m}{\chi}\right)}-6.
\ea
The field equations for a Robertson Walker metric can be taken from ref. \cite{CWCG},
\ba\label{C6b}
&&\left(\frac{2}{\ln\left(\frac{\bar m}{\chi}\right)}-6\right)
\left(\frac{\ddot\chi}{\chi}+3H\frac{\dot\chi}{\chi}\right)
+\ln^{-2}\left(\frac{\bar m}{\chi}\right)\left(\frac{\dot\chi}{\chi}\right)^2\nn\\
&&+\bar\lambda\chi^2\left(4\ln\left(\frac{\bar m}{\chi}\right)-1\right)=12H^2+6\dot H,
\ea
and
\be\label{C6c}
\left(H+\frac{\dot\chi}{\chi}\right)^2=
\frac{1}{3\ln\left(\frac{\bar m}{\chi}\right)}
\left(\frac{\dot\chi}{\chi}\right)^2+\frac{\bar\lambda}{3}\chi^2\ln
\left(\frac{\bar m}{\chi}\right),
\ee
where dots denote derivatives with respect to cosmic time $t$. 

In leading order the solution for $t\to-\infty$ is flat space, such that $\eta=t/\bar a$, with constant scale factor $\bar a$. The scalar field slowly increases with time, according to the implicit expression \cite{CWCG}
\be\label{C6d}
\chi\ln^{\frac12}\left(\frac{\bar m}{\chi}\right)=-\sqrt{\frac{3}{\bar\lambda}}t^{-1}.
\ee
One infers 
\be\label{C6e}
\frac{\dot \chi}{\chi}=-\frac1t~,~H=0,
\ee
and therefore
\be\label{C6f}
\hat\h=-\eta^{-1}.
\ee
This corresponds to the solutions with constant $\nu=0$, discussed in sect. \ref{Graviton correlation}, such that the spectrum of graviton fluctuations is flat, $\nu_T=0$. For the extremely early stages of cosmology the model therefore realizes a flat fluctuation spectrum even though the geometry is Minkowski space. 

In next to leading order $H$ increases slowly 
\be\label{C6g}
H=\frac{\tilde c_H}{\ln^2\left(\frac{\bar m}{\dot \chi}\right)}
\frac{\dot\chi}{\chi}~,~a=a_\infty\exp 
\left\{-\frac{\tilde c_H}{\ln \left(\frac{\bar m}{\chi}\right)}\right\}.
\ee
We can infer $\hat \h$ directly from the field equation \eqref{C6c},
\be\label{C6h}
\hat \h^2=\frac{1}{3\ln\left(\frac{\bar m}{\chi}\right)}
(\p\ln \chi)^2+
\frac{A^2\bar\lambda}{3}\ln 
\left(\frac{\bar m}{\chi}\right).
\ee
This corresponds to the sum of eqs. \eqref{M2-a1} and \eqref{M2-b}. We actually may use eq. \eqref{M2-b} for a direct computation of $\nu$ in eq. \eqref{156}, namely
\ba\label{C6i}
\hat \h^2-\p\hat \h&=&-\nu\hat \h^2=\frac{\hat K}{2}(\p\chi)^2\\
&=&\frac{1}{\ln\left(\frac{\bar m}{\chi}\right)}(\p\ln \chi)^2
=\frac{1}{\ln\left(\frac{\bar m}{\chi}\right)}(\hat \h-\h)^2.\nn
\ea
Within the validity of our approximation $|\h|$ is small as compared to $|\hat \h|$ and can be neglected, such that the expansion for $\nu$ is manifestly frame invariant. One obtains in leading order the tensor spectral index 
\be\label{C6j}
\nu_T=2\nu=-\frac{2}{\ln \left(\frac{\bar m}{\chi}\right)}.
\ee
It vanishes for $\chi\to 0$ and takes slightly negative values for $\chi\ll \bar m$. In the Einstein frame the regime of $\chi\ll\bar m$ corresponds to inflationary cosmology \cite{CWCG}. Inflation ends once $\chi$ reaches values of the order $\bar m$. Our equations are no longer valid in this regime, and we refer for a more complete discussion to ref. \cite{CWCG}. This model provides for a realistic setting of cosmon inflation \cite{VG29}.

\section{Conclusions}
\label{Conclusions}

We have expressed the time evolution of field expectation values and correlation functions in terms of the first and second functional derivatives of the quantum effective action. No information beyond the effective action is needed for these evolution equations. In particular, the exact evolution equation for the correlation functions embodies already the normalization information that is encoded in commutator relations in the operator formalism. On the level of the quantum effective action only ``classical'' fields and correlations appear, and no operators. A discussion of cosmology therefore needs an ansatz(or even better a computation in quantum gravity) for the effective action, plus initial conditions for the values of fields and correlations.

We have expressed the evolution equations for fields and correlation functions in terms of variables that are invariant under conformal scalings of the metric with functions $w(\chi)$ depending on a scalar field. This makes field relativity manifest: all observables have the same values in all frames related by such conformal transformations.

We have described cosmology in terms of a scalar field coupled to the metric. The frame invariant analogue of the scale factor $a$ is the dimensionless ratio between the scale factor and the (variable) Planck length $1/\sqrt{F}$, i.e. $A=a\sqrt{F}$. In primordial cosmology $A(\eta)$ increases  with conformal time $\eta$. In the Einstein frame with fixed Planck mass, $\sqrt{F}=M$, this describes the usual inflationary expansion of the scale factor. In pictures of variable gravity where $\sqrt{F}=\chi$, this expansion of geometry is no longer needed. It can be replaced by an increase of $\chi$. In particular, we have presented an example where a realistic almost scale invariant fluctuation spectrum is generated in a geometry that becomes asymptotically flat Minkowski space. Obviously, a geometric horizon plays no longer a role for the understanding of the time history of fluctuations since $H^{-1}$ tends to infinity.

Our approach demonstrates clearly that the features relevant for an understanding of primordial fluctuations need not to be linked to geometry. What matters is the evolution of dimensionless ratios as $A$. Since the effective action can also arise from a formulation of classical statistical theories, it remains to be seen which features involve genuine quantum effects.

\LARGE
\section*{Appendix A: Second functional derivative for variable gravity}
\renewcommand{\theequation}{A.\arabic{equation}}
\setcounter{equation}{0}

\normalsize

In this appendix we expand the effective action for variable gravity \eqref{166} in second order in the metric fluctuation $h_{\mu\nu}$ and the scalar fluctuation $\delta\chi=\chi-\bar \chi(\eta)$. We write
\be\label{C1}
\Gamma_{2}=\int_x\sqrt{\bar g}(A_V+A_K+A_F),
\ee
with 
\be\label{C2}
A_V=\frac V8(h^2-2h^\nu_\mu h^\mu_\nu)+\frac12\frac{\partial V}{\partial \chi}h\delta \chi+\frac12\frac{\partial^2V}{\partial\chi^2}\delta\chi^2,
\ee
and
\ba\label{C3}
A_K&=&\frac{K}{16}\partial^\mu\bar \chi\partial_\mu\bar \chi(h^2-2h^\nu_\rho h^\rho_\nu)\nn\\
&&+\frac{K}{4}\big[\partial^\mu\bar \chi\partial^\nu\bar \chi(2h_{\mu\rho}h^\rho_\nu-h h_{\mu\nu})\nn\\
&&+2\partial^\mu\bar \chi(h\partial_\mu\delta\chi-2h^\nu_\mu\partial_\nu\delta\chi)+2\partial^\mu\delta\chi\partial_\mu\delta\chi]\nn\\
&&+\frac14\frac{\partial K}{\partial\chi}\partial^\mu\bar \chi\big[\partial^\nu\bar\chi(h\bar g_{\mu\nu}-2h_{\mu\nu})+4\partial_\mu\delta\chi]\delta\chi\nn\\
&&+\frac14\frac{\partial^2K}{\partial\chi^2}\partial^\mu\bar \chi\partial_\mu\bar \chi\delta\chi^2.
\ea

For the piece $A_F$ we need the second variation of the curvature scalar,
\ba\label{B14}
(g^{\frac12}R)_{(2)}&=&\frac12\bar g^{\frac12}
\Big\{\bar R\left(\frac14 h^2-\frac12 h^\rho_\mu h^\mu_\rho\right)-\bar R^{\mu\nu}h h_{\mu\nu}\nn\\
&&+h h^{\mn}{_{;\mu\nu}}-hh;^\mu{_\mu}+2R_{(2)}\Big\},
\ea
with
\ba\label{B14A}
&&R_{(2)}=\bar R^{\mn} h_{\nu\rho}h^\rho_\mu
+h^{\mn}h_{;\mu\nu}+h^\mu_\nu h^\nu_{\mu;}{^\rho}{_\rho}\nn\\
&&\hspace{1.0cm}-h^{\mn}(h^\rho_{\nu;\rho\mu}+h^\rho_{\nu;\mu\rho})-\frac12 h^{\mn}{_{;\rho}} h^\rho_{\nu;\mu}+\frac34 h^\mu_{\nu ;\rho}h^\nu_{\mu;}{^\rho}\nn\\
&&\hspace{1.0cm}-h^{\mu\nu}{_{;\nu}}h^\rho_{\mu;\rho}+h^{\mu\nu}{_;{_\nu}}h_{;\mu}-\frac14 h_;{^\mu}h_{;\mu}\big\}.
\ea
We split $A_F$ into different pieces,
\be\label{C4}
A_F=A_{F1}+A_{F2}+A_{F3},
\ee
where 
\ba\label{C5}
A_{F1}&=&-\frac{F}{8}(h^{\mu\nu}D^2h_{\mu\nu}-hD^2h)\nn\\
&&+\frac{F}{12}\bar R(h^{\mu\nu}h_{\mu\nu}-\frac14h^2)\nn\\
&&-\frac F4\bar C^{\mu\rho\nu\tau}h_{\mu\nu}h_{\rho\tau}.
\ea
The Weyl tensor $\bar C^{\mu\rho\nu\tau}$ vanishes for conformally flat geometries. The piece $A_{F3}$ does not contribute for $h^\nu_{\mu;\nu}=0$,
\be\label{C6}
A_{F3}=\frac F4
(3h^{\mu\nu}h^\rho_{\nu;\rho\mu}+2h^{\mu\nu}{_{;\nu}}h^\rho_{\mu;\rho}-hh^{\mu\nu}{_{;\nu\mu}}).
\ee
Finally, the piece $A_{F2}$ vanishes for constant $F$,
\ba\label{C7}
A_{F2}&=&\frac{\pf}{\pc}\partial^\rho\bar\chi
\left[\frac12 h^\mu_\rho h_{;\mu}-\frac18 hh_{;\rho}+\frac38 h^{\mu\nu}h_{\mu\nu;\rho}\right.\nn\\
&&\qquad\left. -\frac14 h^{\mu\nu}h_{\mu\rho;\nu}\right]\nn\\
&&-\frac12\frac{\pf}{\pc}\left[h^{\mu\nu}{_{;\nu\mu}}-h_;{^\mu}{_\mu}-\bar R^{\mu\nu}h_{\mu\nu}+\frac12\bar Rh\right]\delta \chi\nn\\
&&-\frac14\frac{\partial^2F}{\partial^2\chi^2}\bar R\delta\chi^2.
\ea
For Einstein gravity with a standard scalar kinetic term one has $F=M^2,~K=1$.

Simplifications occur if we concentrate on $h^\nu_{\mu;\nu}=0$, conformally flat geometry, and split $h_{\mu\nu}=\tilde b_{\mu\nu}+h\bar g_{\mu\nu}/4$,
\be\label{C8}
A_{F1}=\frac F8\tilde b^{\mu\nu}
\left(-D^2+\frac{2\bar R}{3}\right)\tilde b_{\mu\nu}+\frac{3F}{32}hD^2h,\\
\ee
and
\ba\label{C9}
A_{F2}&=&\frac{\pf}{\pc}\partial^\rho\bar \chi
\left[\frac38\tilde b^{\mu\nu}\tilde b_{\mu\nu;\rho}-\frac{1}{32}hh_{;\rho}\right]\nn\\
&&+\frac{\pf}{\pc}\bar \chi_;{^\rho}{_\nu}
\left[\frac14 h^{\mu\nu} h_{\mu\rho}-\frac12 h h^\nu_\rho\right]\\
&&+\frac{\pf}{\pc}\left[\frac12 D^2 h+\frac12\bar R^{\mu\nu}h_{\mu\nu}-\frac12\bar Rh\right]\delta\chi\nn\\
&&+\frac{\partial^2F}{\partial\chi^2}
\left[\partial^\rho \bar \chi\partial_\nu\bar \chi
\left(\frac14 h^{\mu\nu}h_{\mu\rho}-\frac12hh^\nu_\rho\right)-\frac14\bar R\delta\chi^2\right].\nn
\ea

\LARGE
\section*{Appendix B: Propagator equation}
\renewcommand{\theequation}{B.\arabic{equation}}
\setcounter{equation}{0}
\normalsize

In this appendix we describe briefly the formal setting for the propagator equation in quantum gravity. For simplicity of notation we restrict the discussion to the metric degree of freedom. The scalar field (as well as other degrees of freedom) can be added in a straightforward way. 

We formulate quantum gravity as a functional integral for the partition function
\be\label{1}
Z[K^{\mu\nu}]=\int\tilde{D}g'_{\rho\sigma}\exp\Big\{-S[g'_{\rho\sigma}]+\int_x g'_{\mu\nu}(x)K^{\mu\nu}(x)\Big\}.
\ee
The regularization of this functional integral as, for example, gauge fixing and ghost terms, are here formally included in the functional measure $\int\tilde{D}g'_{\rho\sigma}$. The action $S$ is supposed to be invariant under general coordinate transformations or diffeomorphisms.
The source $K^{\mu\nu}=K^{\nu\mu}$ transforms as a contravariant tensor density. For the functional measure we will employ a background field formalism such that the measure is variant under a simultaneous diffeomorphism transformation of the background metric and the fluctuations, see below. Therefore $Z$ is invariant under this combined transformation. The background metric $\bar g_{\mu\nu}$ is also used to relate the source $K^{\mu\nu}$ to the energy momentum tensor $T^{\mu\nu}$,
\be\label{8}
K^{\mu\nu}=\frac{1}{2}{\bar{g}}^{\frac12}T^{\mu\nu}, \quad \bar g=\det(\bar{g}_{\mu\nu}).
\ee

As usual, the quantum effective action obtains from $W(K)$ by a Legendre transform
\be\label{L1}
\Gamma[g_{\mn}]=-W[K^{\mn}]+\int_xg_{\mn}K^{\mn}.
\ee
It is well defined if the expectation value of the metric
\be\label{L2}
g_{\mn}(x)=\kl g'_{\mn}(x)\kr=
\frac{\delta W}{\delta K^{\mn}(x)}
\ee
is uniquely defined for a given source $K^{\mn}$, such that eq. \eqref{L2} can be inverted and $K^{\mn}[g_{\rho \sigma}]$ inserted in eq. \eqref{L1}. The invertibility of eq. \eqref{L2} can be realized in different ways. Either one adds a gauge fixing term to action, e.g. 
\be\label{L3}
S_{gf}=\frac{1}{2\beta}\int_x\sqrt{\bar g}h'^\nu_{\mu;\nu}h'^{\mu\rho}{_{;\rho}}~,~
h'_{\mn}=g'_{\mn}-\bar g_{\mn},
\ee
with covariant derivatives denoted by semicolons involving the connection for the background metric $\bar g_{\mn}$. For $\beta\to\infty$ this imposes on $g_{\mn}$ the constraint
\be\label{L4}
h^\nu_{\mu;\nu}=0~,~g_{\mn}=\bar g_{\mn}+h_{\mn}.
\ee
Alternatively, we may restrict the source terms to covariantly conserved energy momentum tensors $T^\nu_{\mu;\nu}=0$. As a result $g_{\mn}$ is a constrained field obeying eq. \eqref{L4}.

We may interpret the second functional derivatives $\Gamma^{(2)}$ and $W^{(2)}$ as matrices. They obey the usual matrix identity
\ba\label{29}
\Gamma^{(2)}W^{(2)}&=&1, \\
\int_y\Gamma^{(2)\mn\rho\tau}(x,y)W^{(2)}_{\rho\tau\sigma\lambda}(y,z)&=&
E^{\mn}{_{\sigma\lambda}}(x,z),\nn
\ea
that follows directly from the defining relations for $\Gamma$. 
Here $E^{\mn}{_{\sigma\lambda}}$ is the unit matrix in the space of appropriate functions. In the gauge fixed version with unconstrained $h_{\sigma\lambda}$ the unit matrix reads $E^{\mn}{_{\sigma\lambda}}=\frac12(\delta^\mu_\sigma\delta^\nu_\lambda+\delta^\mu_\lambda\delta^\nu_\sigma)\delta(x-z)$, while in the presence of a constraint \eqref{L4} it becomes a projector on the subspace of fields obeying eq. \eqref{L4}. 

On the other hand, $W^{(2)}$ defines the connected two-point correlation function (Green's function, propagator)
\ba\label{30}
W^{(2)}_{\rho\tau\sigma\lambda}(x,y)&=&\langle h_{\rho\tau}(x)h_{\sigma\lambda}(y)\rangle_c\\
&=&\langle h_{\rho\tau}(x)h_{\sigma\lambda}(y)\rangle - \langle h_{\rho\tau}(x)\rangle \langle h_{\sigma\lambda}(y)\rangle.\nn
\ea
Eq.\eqref{29} is therefore an exact "propagator equation" for the Green's function
\be\label{37AA}
G_{\rho\tau\sigma\lambda}(x,y)=W^{(2)}_{\rho\tau\sigma\lambda}(x,y).
\ee

If $\Gamma^{(2)}$ contains time-derivatives this is an evolution equation which describes the time dependence of the Green's function. Typically, $\Gamma^{(2)}$ is of the form
\be\label{37AB}
\Gamma^{(2)\mu\nu\rho\tau}(x,y)=\delta(x-y)\Gamma^{(2)\mu\nu\rho\tau}(y),
\ee
where $\Gamma^{(2)\mu\nu\rho\tau}(y)$ contains derivatives with respect to $y$. The resulting propagator equation reads
\be\label{37AC}
\Gamma^{(2)\mu\nu\rho\tau}(x)G_{\rho\tau\sigma\lambda}(x,y)=
E^{\mn}{_{\sigma\lambda}} (x,y).
\ee

By construction, the effective action depends separately on $g_{\mn}$ and $\bar g_{\mn}$. We may write
\be\label{L5}
\Gamma [g_{\mn},\bar g_{\mn}]=\Gamma[g_{\mn}]+\Gamma_{gf}
[h_{\mn},\bar g_{\mn}],
\ee
where the first term identifies both metrics,
\be\label{L6}
\Gamma[g_{\mn}]=\Gamma[g_{\mn},\bar g_{\mn}=g_{\mn}].
\ee
In a gauge fixed version the second term contains, in particular, the gauge fixing term. We will work in the approximation where $\Gamma_{gf}$ solely acts as a generalized gauge fixing on the level of the effective action, in the sense that its contribution to the effective action for the gauge invariant fluctuations (Bardeen potentials) $\Phi, \Psi,\Omega_m$ and $\gamma_{mn}$ can be omitted.

\LARGE
\section*{Appendix C:\newline Graviton fluctuations for variable gravity}
\renewcommand{\theequation}{C.\arabic{equation}}
\setcounter{equation}{0}
\normalsize

For the computation of the effective action in second order in the graviton fluctuations we start with metric fluctuations that are traceless and divergence free
\be\label{A16-1}
h_{\mn}=t_{\mn}~,~t^\nu_{\mu;\nu}=0~,~t^\mu_\mu=0.
\ee
This expression is inserted into $\Gamma_2$ as displayed in appendix A. 
Typical terms in $\Gamma^{(2)}$ involve the covariant Laplacian, as
\be\label{A16A}
\Gamma^{(2)\mu\nu\rho\tau}_{tt}=-\frac{F}{4}\sqrt{\bar g}P^{(t)\mu\nu\rho\tau}
\left(D^2-\frac{\bar R}{6}\right)+\dots,
\ee
with $P^{(t)}$ an appropriate projector. We need the covariant Laplacian acting on a traceless tensor $b_{\mn}$,
\ba\label{P2}
D^2 b_{00}&=&-\frac{1}{a^2}\Big\{ \big(\partial^2_\eta-2{\cal H}\partial_\eta-2\partial_\eta{\cal H}-8{\cal H}^2+k^2\big)b_{00}\nn\\
&&+4i{\cal H}\delta^{jl}k_jb_{l0}\Big\},\nn\\
D^2 b_{m0}&=&-\frac{1}{a^2}\Big\{\big(\partial^2_\eta-2{\cal H}\partial_\eta-2\partial_\eta{\cal H}-6{\cal H}^2+k^2)b_{m0}\nn\\
&&+2i{\cal H}k_mb_{00}+2i{\cal H}\delta^{jl}k_j b_{ml}\Big\},\nn\\
D^2 b_{mn}&=&-\frac{1}{a^2}\Big\{(\partial^2_\eta-2{\cal H}\partial_\eta-2\partial_\eta{\cal H}-2{\cal H}^2+k^2)b_{mn}\nn\\
&&-2{\cal H}^2\delta_{mn}b_{00}
+2{\cal H}(ik_mb_{n0}+ik_nb_{m0})\Big\},
\ea
such that the operator $D^2$ in eq. \eqref{A16A} reads 
\be\label{S5}
D^2=-\frac{1}{a^2(\eta)}
(\partial^2_\eta-2{\cal H}\partial_\eta -2\partial_\eta{\cal H}-2{\cal H}^2+k^2).
\ee

Restricting next $t_{mn}=a^2\gamma_{mn},~t_{m0}=0,~t_{00}=0$ one arrives at 
\ba\label{167}
\Gamma_\gamma&=&\int_x\frac{ia^4}{8} \gamma^{mn*}
\Big\{a^2F(\partial^2_\eta+2\h\partial_\eta\nn\\
&&-3\partial_\eta\ln F\partial_\eta+4\partial_\eta\h+2\h^2-2\h\partial_\eta\ln F+k^2)\nn\\
&&-2a^4V+a^2 K(\p\bar \chi)^2\Big\}\gamma_{mn}.
\ea
The terms $\sim \p F$ arise from partial integration of terms in $R_{(2)}$ for which derivatives act on both $h_{\mu\nu}$-factors. For $h^\nu_{\mu;\nu}=0$ the part responsible for the terms involving $\p F$ reads
\be\label{168}
\Delta\Gamma_2=\int_x\sqrt{\bar g}\partial_\rho F
\left\{\frac38 h^{\mu\nu}h_{\mu\nu;}{^\rho}-\frac14h^{\mu\nu}h^\rho_{\nu;\mu}-\frac18 h h_;^\rho\right\}.
\ee

Expressing eq. \eqref{167} in terms of $A,\hat\h,\hat V$ and $\hat K$ yields eq. \eqref{J1}. The first functional derivative is given by 
\ba\label{169}
\frac{\partial\Gamma_\gamma}{\partial\gamma^*_{mn}}&=&\frac i4
P^{(\gamma)mnm'n'}
\Big[A^2\big\{\partial^2_\eta+2\hat \h\partial_\eta+4\partial_\eta\hat\h+2\hat \h^2+k^2\big\}\nn\\
&&-2A^4\hat V+A^2\hat K(\partial_\eta\bar\chi)^2\Big]\gamma_{m'n'},\nn\\
\ea
with $P^{(\gamma)}$ given by eq. \eqref{159A}. For the second functional derivative one omits $\gamma_{m'n'}$ on the r.h.s. of eq. \eqref{169}.

\LARGE
\section*{Appendix D: Vector fluctuations for variable gravity}
\renewcommand{\theequation}{D.\arabic{equation}}
\setcounter{equation}{0}

\normalsize

In this appendix we expand the effective action \eqref{166} of variable gravity to second order in the vector fluctuations. For this purpose we insert eq. \eqref{Z8} in the expression $\Gamma_2$ displayed in appendix A. In order to keep oversight we discuss different pieces separately. The trace $h$ does not contribute to the vector fluctuations, such that it is sufficient to consider the traceless part $\tilde b_{\mn}$ of $h_{\mn}$. 

We write
\be\label{Z11}
\Gamma^{(V)}_{2}=\int_{\eta,k}ia^4(B_{V1}+B_{V2}+B_{V3}+B_{V4}).
\ee
The first part arises from $A_{F1}$ in eq. \eqref{C5},
\ba\label{Z12}
B_{V1}&=&-\frac{F}{8}
(\tilde b^{mn*}D^2\tilde b_{mn}+2\tilde b^{m0*}D^2\tilde b_{m0}).
\ea
Expressing $D^2$ by eq. \eqref{P2} yields 
\ba\label{Z15}
B_{V1}&=&-\frac F4\Big\{W^{m*}(\partial^2_\eta+2\h\partial_\eta-6\h^2+k^2)W_m\nn\\
&-&k^2 V^{m*}(\partial^2_\eta+2\h\partial_\eta-2\h^2+k^2)V_m\nn\\
&&- 2\h k^2(W^{m*}V_m+V^{m*}W_m)\Big\}.
\ea
For the second term we combine contributions from $A_V,~A_K$ and $A_{F1}$, namely
\ba\label{Z12A}
B_{V2}&=&\left(\frac{F\overline{R}}{12}-\frac{V}{4}-\frac{K}{8}\partial^{\mu}\overline{\chi}\partial_{\mu}\overline{\chi}\right)(\tilde{b}^{mn*}\tilde{b}_{mn}+2\tilde{b}^{m0*}\tilde{b}_{m0})\nn\\
&&\qquad\qquad-\frac{K}{2a^2}(\partial_{\eta}\bar{\chi})^2\tilde{b}^{m0*}\tilde{b}_{m0}\nn\\
&=&\frac{a^2}{2}\left(V-\frac{F\bar R}{3}\right)
(W^{m*}W_m-k^2V^{m*}V_m)\\
&&\qquad\qquad+\frac K4(\partial_\eta\bar\chi)^2
(W^{m*}W_m+k^2V^{m*}V_m).\nn
\ea

The part $B_{V3}$ arises from $A_{F3}$ in eq. \eqref{C6} and involves
\be\label{XABC}
Z_{\mu}=h^{\nu}_{\mu;\nu}.
\ee
It vanishes for covariantly conserved metric fluctuations,
\ba\label{Z12B}
&&B_{V3}=\frac{F}{4}(3\tilde{b}^{\mu\nu*}Z_{\nu;\mu}+2Z^{\mu*}Z_{\mu})\nn\\
&&=\frac{F}{4} \Big\{3W^{m*}\big[(\partial^2_\eta+2\h\partial_\eta+4\partial_\eta\h-8\h^2)W_m\\
&&+(\partial_\eta-2\h)k^2V_m\big] -3k^2V^{m*}\big[(\partial_\eta+4\h)W_m+k^2V_m\big]\nn\\
&&+2\big[(\partial_\eta+6\h)W^{m*}+k^2V^{m*}\big]
\big[(\partial_\eta+4\h)W_m+k^2V_m\big]\Big\}.\nn
\ea
Finally, $B_{V4}$ vanishes for constant $F$, 
with $\big(\partial_\eta F=(\partial F/\partial\bar \chi)\partial_\eta\bar\chi$,
\ba\label{Z12C}
B_{V4}&=&-\frac{\partial_\eta F}{a^2}
\left\{\frac38\tilde b^{mn*}\tilde b_{mn;0}-\frac14\tilde b^{mn*}\tilde b_{m0;n}\right.\nn\\
&&+\left.\frac12\tilde b^{m0*}\tilde b_{m0;0}-\frac14\tilde b^{m0*}
\tilde b_{00;m}\right\}\\
&=&\frac14\p F\big\{2W^{m*}(\p+\h)W_m\nn\\
&&+k^2(V^{m*}W_m-2\h V^{m*}V_m-3V^{m*}\p V_m)\big\}.\nn
\ea

Let us next group the different terms. By partial integration one finds 
\ba\label{Z15A}
&&B_{V1}+B_{V3}=\frac F4\Big\{ W^{m*}(4\partial_\eta\h-2\h^2-k^2)W_m\nn\\
&&+k^2W^{m*}\partial_\eta V_m
-k^2V^{m*}(\partial_\eta+2\h)W_m\\
&&+k^2V^{m*}
(\partial^2_\eta+2\h\partial_\eta-2\h^2)V_m\Big\}+B_{V5},\nn
\ea
with $B_{V5}$ vanishing for constant $F$,
\ba\label{Z15Aa}
&&B_{V5}=-\frac12\partial_\eta F W^{m*}\big[(\partial_\eta+4\h)W_m+k^2V_m\big]\nn\\
&&\qquad +\frac14\p FV^{m*}(W_m-\p V_m).
\ea
In terms of the ``gauge invariant'' fluctuation $\Omega_m$,
\be\label{Z10}
\partial_\eta V_m=W_m-\Omega_m,
\ee
we obtain
\ba\label{301A}
&&B_{V1}+B_{V3}=-\frac F4\{k^2\Omega^{m*}\Omega_m+2\h^2k^2V^{m*}V_m\nn\\
&&\qquad+(2\h^2-4\partial_\eta\h)W^{m*}W_m\}+B_{V5}.
\ea

We next write
\ba\label{301B}
B_{V2}&=&\frac F2
\big\{\h^2k^2V^{m*}V_m+(\h^2-2\partial_\eta\h)W^{m*}W_m\big\}\nn\\
&&+B_{V6}+B_{V7},
\ea
where $B_{V6}$ vanishes again for constant $F$,
\be\label{301C}
B_{V6}=\frac{3\h}{2}\partial_\eta FW^{m*}W_m-\frac{k^2}{2}
(\partial^2_\eta F+\h\partial_\eta F)V^{m*}V_m.
\ee
The contribution
\ba\label{301D}
B_{V7}&=&-\frac14
\Big\{2a^2V-K(\partial_\eta\bar \chi)^2-F
\big[4\partial_\eta\hat \h+2\hat \h^2\nn\\
&&+\frac32(\partial_\eta\ln F)^2\big]\Big\}
[W^{m*}W_m+k^2V^{m*}V_m]\nn\\
&&+\big\{a^2V-F(\partial_\eta\hat\h+2\hat \h^2)\big\}W^{m*}W_m
\ea
vanishes if the field equations \eqref{M2-a1}, \eqref{M2-b} are obeyed.

Taking things together yields
\be\label{301E}
\Gamma^{(V)}_2=-\frac i4\int_{\eta,k}a^4Fk^2\Omega^{m*}\Omega_m+i\int\limits_{\eta,k}a^4\Delta B_V
\ee
where
\be\label{301F}
\Delta B_V=B_{V4}+B_{V5}+B_{V6}+B_{V7}.
\ee
The combination $B_{V4}+B_{V5}+B_{V6}$ involves time derivatives of $F$ and vanishes,
\ba\label{301Fa}
B_{V4}+B_{V5}+B_{V6}&=&-\frac{k^2}{2}V^{m*}(\partial^2_\eta F+2\h\p F\nn\\
&&+2\p F\p)V_m\\
&=&-\frac{k^2}{2a^4}\p
(a^4\p FV^{m*}V_m)=0.\nn
\ea

~

\vspace{-0.1cm}\noindent
Here we employ the property that this expression stands under a $k$-integral such that we can use effectively $W^{m*}V_m=V^{m*}W_m$. The last identity uses the fact that multiplication of $B_{V4}+B_{B5}+B_{V6}$ by $a^4$ yields a total derivative. We conclude that $\Delta B_V$ is given by $B_{V7}$ and vanishes if the background obeys the field equations. Using frame invariant variables, and $\Omega^m=a^{-2}\Omega_m$ etc., eq. \eqref{301E} yields eq. \eqref{A20-1}. 

This somewhat lengthly computation can be simplified for the case of covariantly conserved metric fluctuations. We may impose $h^\nu_{\mu;\nu}=0$ and express $\partial_\eta V_m$ in terms of the gauge invariant vector $\Omega_m$. The relation $h^\nu_{\mu;\nu}=0$ reads
\be\label{Z9}
\partial_\eta W_m =-k^2V_m-4\h W_m.
\ee 
We can then employ eqs. \eqref{Z9}, \eqref{Z10} in order to eliminate the $\eta$-derivatives acting on $V_m$ or $W_m$, such that $B_{V3}=0$ and 
\ba\label{Z16}
B_{V1}&=&-\frac F4 \Big\{k^2V^{m*}\big[(\partial_\eta+2\h)\Omega_m+2\h^2 V_m\big]\\
&&+k^2W^{m*}\Omega_m-(4\partial_\eta\h-2\h^2)W^{m*}W_m\Big\}.\nn
\ea
Using partial integration we can replace $k^2 V^{m*}(\partial_\eta+2\h)\Omega_m$ by $k^2\Omega^{m*}\Omega_m-k^2 W^{m*}\Omega_m-\partial_\eta\ln FV^{m*}\Omega_m$, such that 
\ba\label{Z17}
B_{V1}&=&-\frac F4\Big\{ k^2\Omega^{m*}\Omega_m+2\h^2k^2V^{m*}V_m\nn\\
&&-(4\partial_\eta\h-2\h^2)W^{m*} W_m\Big\}+B_{V5},
\ea
with 
\be\label{Z18}
B_{V5}=\frac14\frac{\pf}{\pc}\partial_\eta\bar\chi V^{m*}\Omega_m.
\ee
If we impose the solution of the field equation for the background metric the second order expansion of the effective action in the vector channel becomes
\be\label{Z19}
\Gamma^{(V)}_{2}=-\int \frac{ia^4Fk^2}{4}\Omega^{m*}\Omega_m.
\ee


\bibliography{Primordial_fluctuations_for_variable_gravity}

\end{document}